\documentclass[12pt]{article}

\usepackage{epsfig}

\begin{document}

\begin{center}
{\Large \bf Top quark and charged Higgs production at hadron colliders}
\end{center}
\vspace{2mm}
\begin{center}
{\large Nikolaos Kidonakis}\\
\vspace{2mm}
{\it Kennesaw State University, Physics  \#1202\\
1000 Chastain Rd., Kennesaw, GA 30144-5591, USA}\\
\end{center}

\begin{abstract}
I present a brief theoretical update on top quark pair production at the 
Tevatron and give values of the NNLO-NNNLL cross section 
for both $m_t=175$ and 178 GeV.
I then present a calculation of the cross section for charged Higgs 
production in association with a top quark at the LHC, including
NNLO soft-gluon corrections.
\end{abstract}

\section{Top quark production at the Tevatron}

The properties of the top quark, in particular its mass 
and production cross section, are subjects of intense study at
the Tevatron \cite{CDF,D0}. 
The most accurate theoretical prediction \cite{NKRV} for top quark pair 
production at the Tevatron includes soft-gluon corrections 
\cite{KS,NKhq,NKuni}
through next-to-next-to-next-to-leading logarithmic (NNNLL)
accuracy at next-to-next-to-leading order (NNLO), denoted
as NNLO-NNNLL \cite{NKRV}.
These corrections are sizable and provide a dramatic decrease
in the scale dependence of the cross section. Results have been
derived in both single-particle-inclusive (1PI) kinematics and 
pair-invariant-mass (PIM) kinematics. There are differences in the 
results in the two kinematics due to subleading terms, and the best 
estimate is given by the average of the two kinematics.
  
For a top quark mass $m_t=175$ GeV the theoretical value of the
cross section is \cite{NKRV}
\newline

$\sigma_{t \bar t}^{NNLO-NNNLL}$($\sqrt{S}=1.8$ TeV, $m_t$=175 GeV)
$=5.24 \pm 0.31$ pb \hspace{3mm} and 
\newline

$\sigma_{t \bar t}^{NNLO-NNNLL}$($\sqrt{S}=1.96$ TeV, $m_t$=175 GeV)
$=6.77 \pm 0.42$ pb
\newline

\noindent at Run I and Run II, respectively.
The uncertainty indicated is due to the kinematics ambiguity;
the scale uncertainty is much smaller.

Some recent data from the Tevatron suggest a value for the top quark mass
around  $m_t=178$ GeV. For that value of top mass the theoretical
cross sections become
\newline

$\sigma_{t \bar t}^{NNLO-NNNLL}$($\sqrt{S}=1.8$ TeV, $m_t$=178 GeV)
$=4.76 \pm 0.28$ pb \hspace{3mm} and
\newline

$\sigma_{t \bar t}^{NNLO-NNNLL}$($\sqrt{S}=1.96$ TeV, $m_t$=178 GeV)
$=6.15 \pm 0.38$ pb.
\newline

Results for the top quark transverse momentum distributions 
at NNLO-NNNLL are also available \cite{NKRV}.

\section{Charged Higgs production via $bg \rightarrow t H^-$}

A future discovery of a charged Higgs boson would be an
umistakable sign of new physics beyond the Standard Model \cite{HiggsLH}.
The LHC has good potential for such a discovery  through the
partonic process $bg \rightarrow t H^-$.
The Born cross section is proportional to $\alpha \alpha_s
(m_b^2\tan^2 \beta+m_t^2 \cot^2 \beta)$, where
$\tan \beta=v_2/v_1$ is the  
ratio of the vacuum expectation values (vev's) of two Higgs doublets in 
the MSSM.

Full NLO calculations have recently become available \cite{Zhu,Plehn}, 
and they show that
the NLO corrections are big. Since charged Higgs production will
be a near-threshold process at the LHC, given the expected large
mass of this particle (hundreds of GeV), threshold soft-gluon
corrections can provide significant enhancements of the cross section.
A  next-to-leading logarithm (NLL) calculation of these corrections
at NNLO, denoted as NNLO-NLL \cite{NKch}, showed that indeed the
soft-gluon corrections are substantial and they decrease the scale
dependence of the cross section, thus providing a better theoretical 
prediction.

For the process
$b(p_b) + g(p_g) \longrightarrow t(p_t)+H^-(p_{H^-})$ we
define $s=(p_b+p_g)^2$, $t=(p_b-p_t)^2$, $u=(p_g-p_t)^2$, 
and $s_4=s+t+u-m_t^2-{m_{H^-}}^2$.
At threshold $s_4 \rightarrow 0$.
The soft-gluon corrections take the form   
{$[(\ln^l(s_4/m_{H^-}^2))/s_4]_+$}.
For the order $\alpha_s^n$ corrections, $l \le 2n-1$. The
leading logarithms (LL) are those with 
$l=2n-1$, while for the NLL $l=2n-2$.
We calculate NLO and NNLO corrections at NLL accuracy.

\begin{figure}
\begin{center}
\includegraphics[width=10cm]{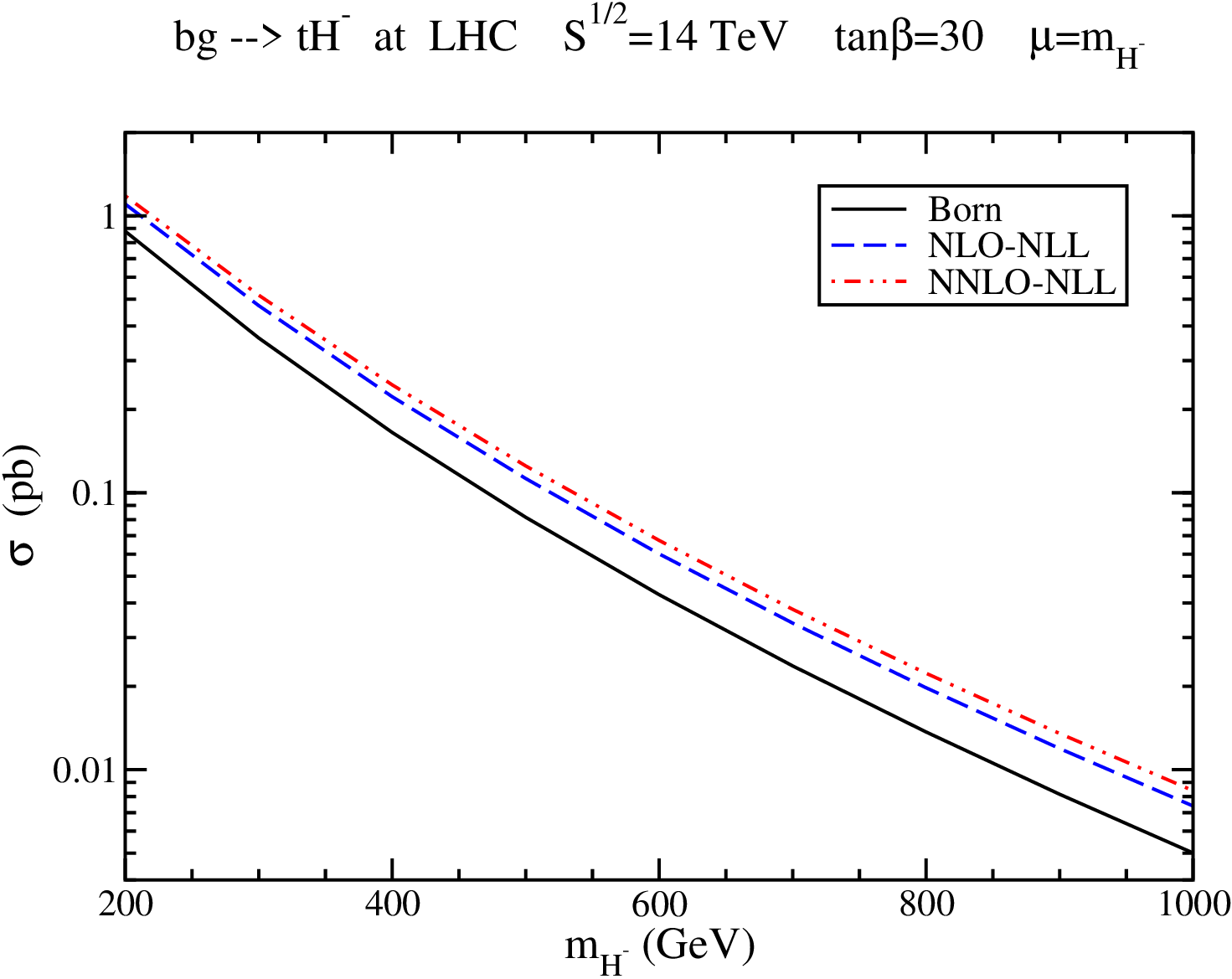} 
\caption{The total cross section for charged Higgs production at the LHC.}
\end{center}
\end{figure}

In Figure 1 we plot the cross section versus
charged Higgs mass for $pp$ collisions at the LHC with $\sqrt{S}=14$ TeV. 
We use the MRST2002 approximate 
NNLO parton distributions functions (PDF) \cite{mrst2002}
with the respective  three-loop evaluation of $\alpha_s$. 
We set the factorization scale equal to the renormalization scale 
and denote this common scale by $\mu$.
We show results for the Born, NLO-NLL, and NNLO-NLL
cross sections, all with a choice of scale $\mu=m_{H^-}$. 
In our calculations we use $\tan \beta=30$. 
The NLO and NNLO threshold corrections are positive and provide a significant
enhancement to the lowest-order result.
We note that the cross sections for the related process
${\bar b} g \rightarrow {\bar t} H^+$ are exactly the same.

\begin{figure}
\begin{center}
\includegraphics[width=6.5cm]{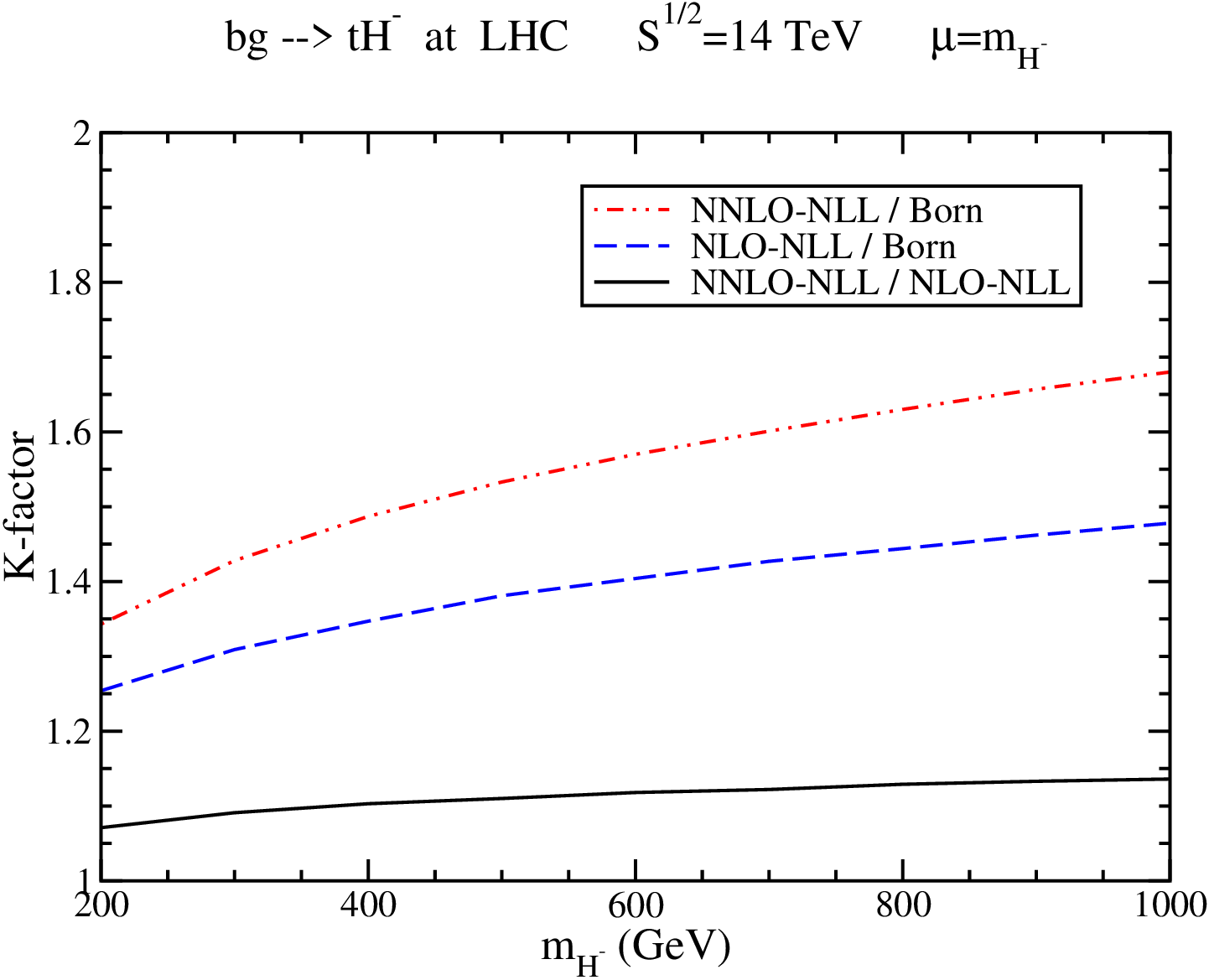} 
\hspace{1mm}
\includegraphics[width=6.5cm]{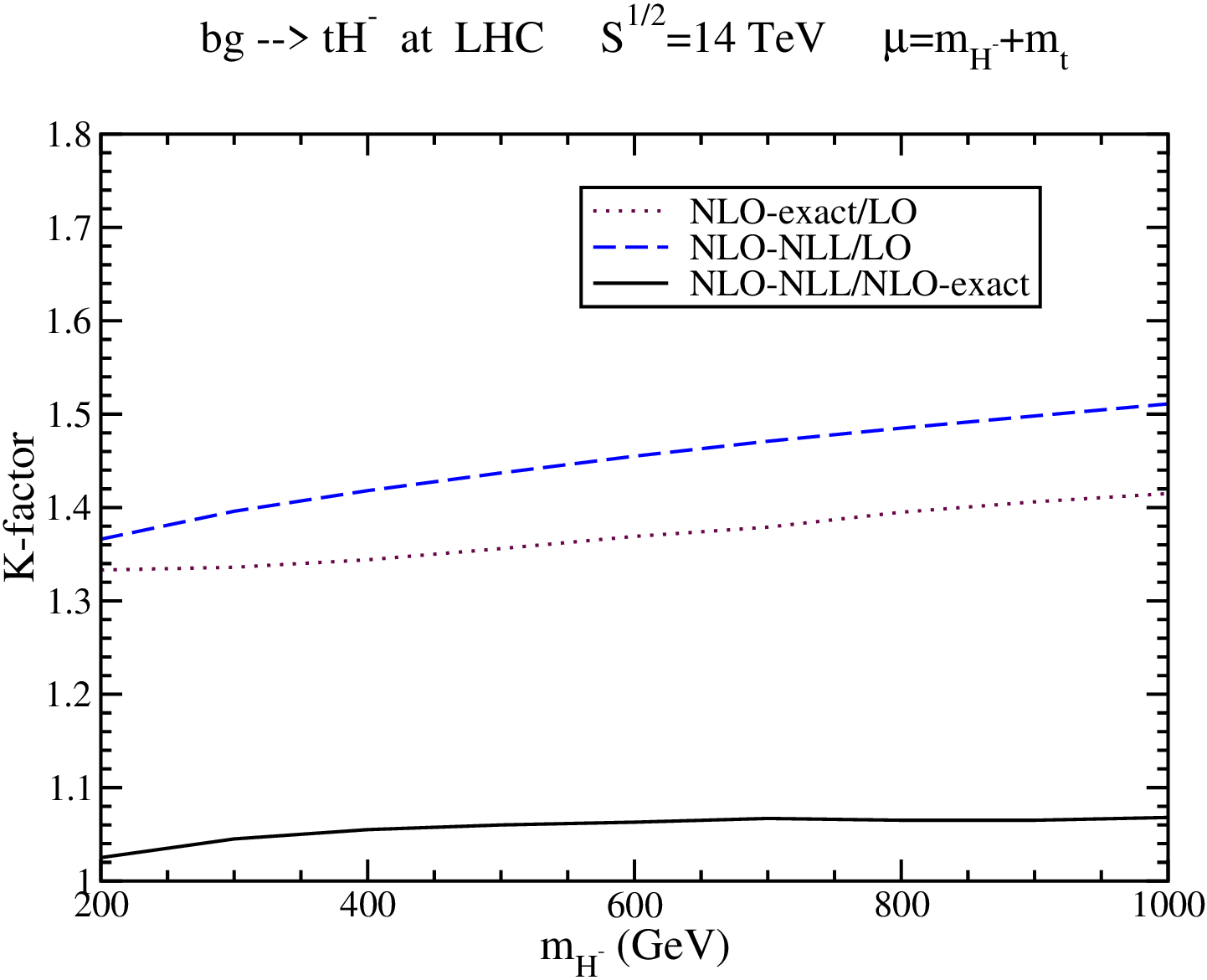}
\caption{$K$-factors for charged Higgs production at the LHC.}
\end{center}
\end{figure}

In Figure 2 we plot $K$-factors, i.e. ratios of cross
sections at various orders. 
On the left-hand side, the NLO-NLL / Born curve shows that the
NLO threshold corrections enhance the Born cross section by approximately
25\% to 50\% depending on the mass of the charged Higgs. 
The NNLO-NLL / Born curve shows that if we include the NNLO threshold 
corrections we get an enhancement over the Born result of approximately
35\% to 70\% in the range of masses shown. 
Finally, the NNLO-NLL / NLO-NLL curve shows clearly the further enhancement 
over NLO that the NNLO threshold corrections provide, between 7\% and
14\%. 
On the right-hand side  we compare our NLO-NLL results with the exact results
that have been derived in \cite{Zhu}. 
To make the comparison with \cite{Zhu}, the NLO-NLL result is calculated 
here for $\mu=m_{H^-}+m_t$, the choice of scale used in that reference,
and also using a two-loop $\alpha_s$. Also the use of $K$-factors removes
any  discrepancies
arising from different choices of parton distribution functions.
The NLO-NLL / NLO-exact curve 
is very close to 1 (only a few percent difference), and this shows 
that the NLO-NLL cross section is a 
remarkably good approximation to the exact NLO result. As noted before,
we might have expected this on theoretical grounds since this is 
near-threshold production, and also from prior experience with many
other near-threshold hard-scattering cross sections
\cite{NKRV,NKhq,HSvar}.

\begin{figure}
\begin{center}
\includegraphics[width=10cm]{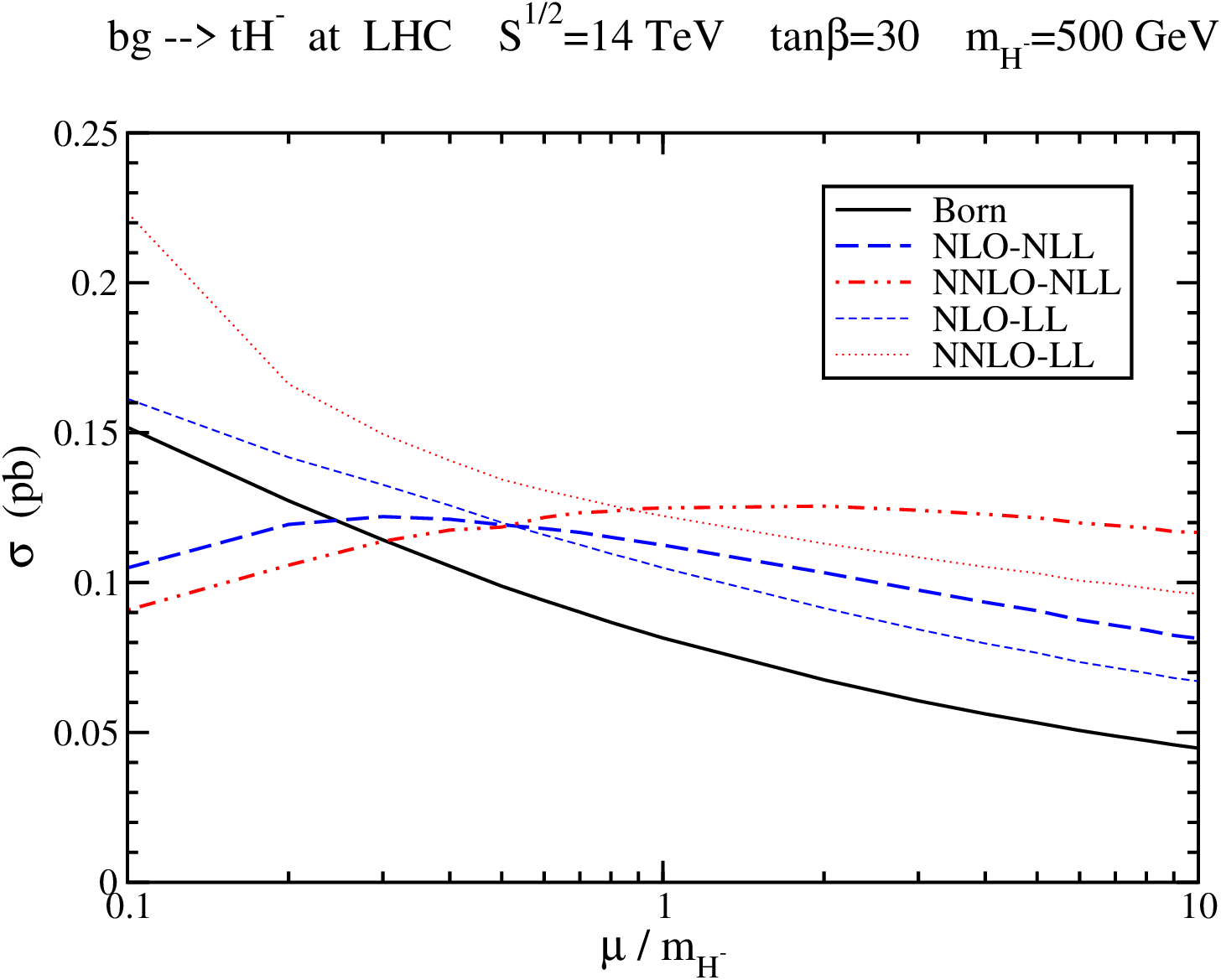} 
\caption{The scale dependence of the charged Higgs cross section.}
\end{center}
\end{figure}

In Figure 3, we plot the scale dependence of the
cross section for a fixed charged Higgs mass $m_{H^-}=500$ GeV.
We plot a large range in scale, $0.1 \le \mu/m_{H^-} \le 10$, and see indeed
that the threshold corrections greatly decrease the scale dependence 
of the cross section. The NNLO-NLL curve is relatively flat.
For comparison we also plot the results using only a leading logarithm (LL)
approximation.
We see that the LL results 
display a large scale dependence at both NLO and NNLO,
and are not an improvement over the Born result. The NLL terms are
essential in diminishing the scale dependence. 
The difference between the LL and NLL results at both NLO
and NNLO can be very substantial. Thus having a complete NLL calculation,
as provided here, is crucial in providing stable theoretical 
predictions.

Finally, we note that even higher-order corrections may provide
sizable contributions to hard-scattering cross sections. In particular
current calculations of next-to-next-to-next-to-leading order
(NNNLO) soft-gluon corrections indicate a non-negligible enhancement 
of the cross section for charged Higgs production.

\end{document}